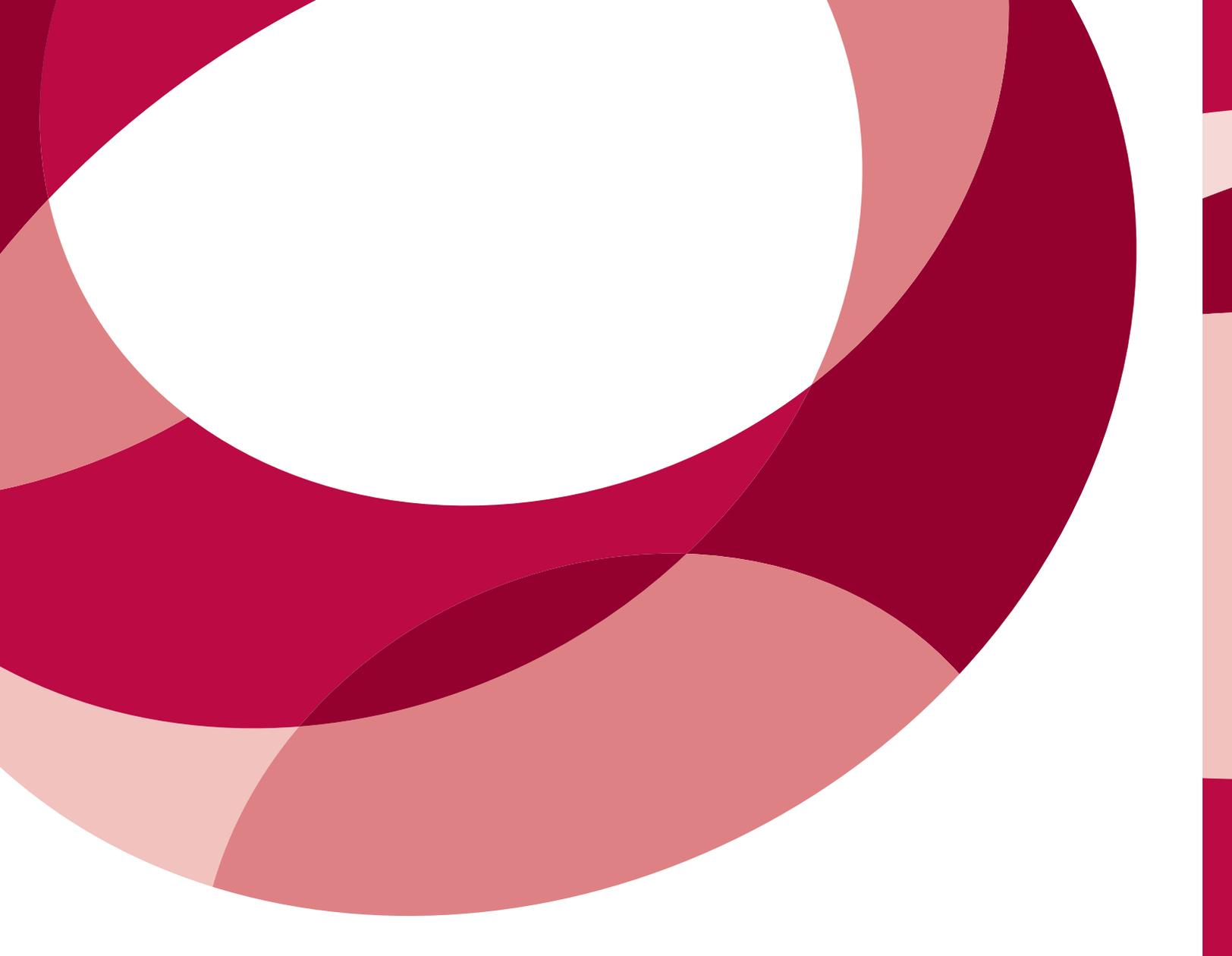

# Identifying Research Challenges in Post Quantum Cryptography Migration and Cryptographic Agility

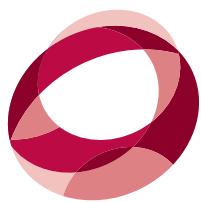

CCC
Computing Community Consortium
Catalyst

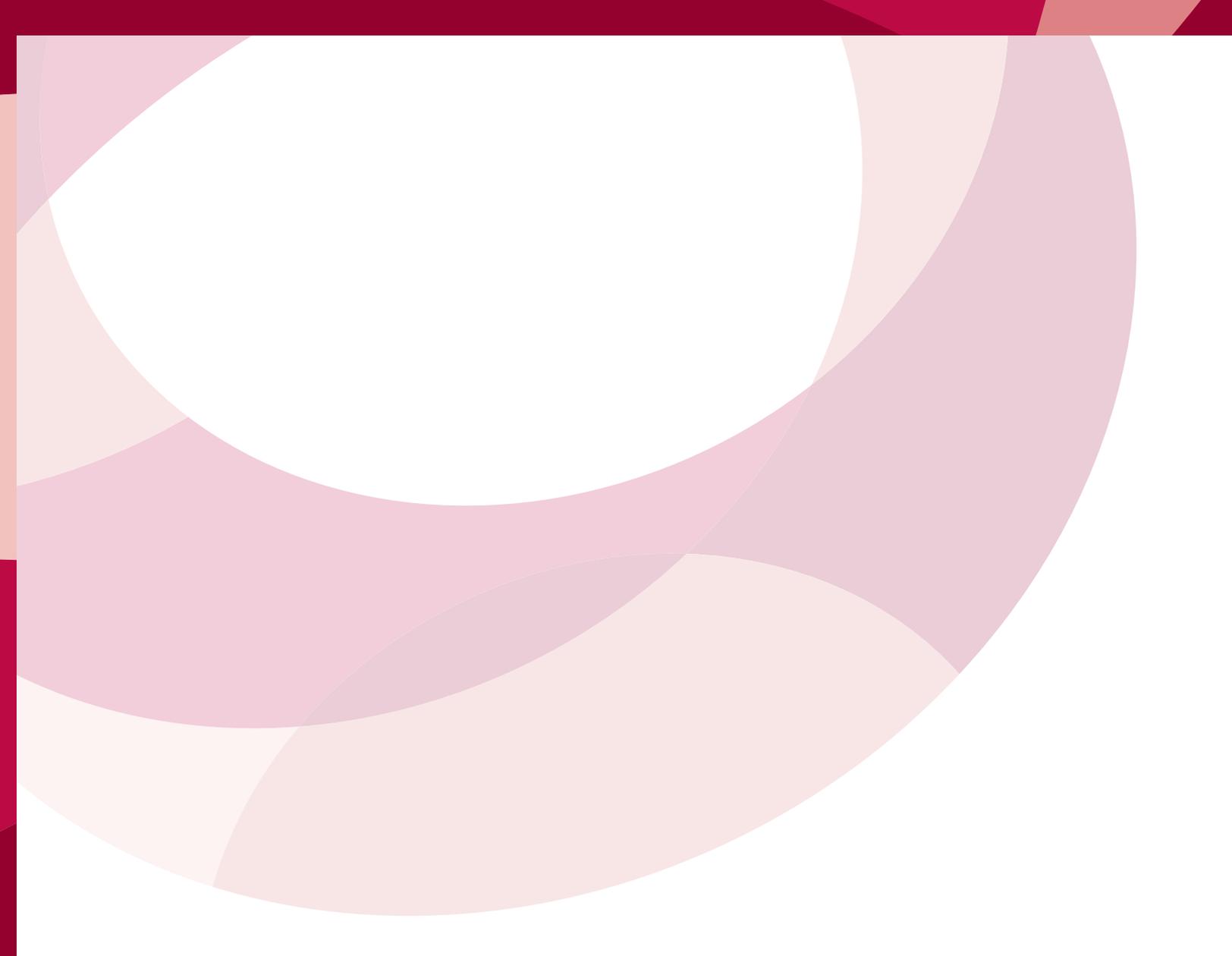

This material is based upon work supported by the National Science Foundation under Grant No. 1734706. Any opinions, findings, and conclusions or recommendations expressed in this material are those of the authors and do not necessarily reflect the views of the National Science Foundation.

# Identifying Research Challenges in Post Quantum Cryptography Migration and Cryptographic Agility

**Organizing Committee:**

David Ott, VMware Research

Christopher Peikert, University of Michigan

**With Support From:**

Mark Hill, University of Wisconsin, Madison and CCC Chair

Ann Schwartz Drobnis, CCC Director

Chris Ramming, VMware

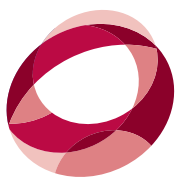

CCC
Computing Community Consortium
Catalyst

## Executive Summary

The implications of sufficiently large quantum computers for widely used public-key cryptography is well-documented and increasingly discussed by the security community. An April 2016 report by the National Institute of Standards and Technology (NIST), notably, calls out the need for new standards to replace cryptosystems based on integer factorization and discrete logarithm problems, which have been shown to be vulnerable to Shor's quantum algorithm for prime factorization. Specifically, widely used RSA, ECDSA, ECDH, and DSA cryptosystems will need to be replaced by **post-quantum cryptography (PQC)** alternatives (also known as *quantum-resistant* or *quantum-safe* cryptography). Failure to transition before sufficiently powerful quantum computers are realized will jeopardize the security of public key cryptosystems which are widely deployed within communication protocols, digital signing mechanisms, authentication frameworks, and more. To avoid this, NIST has actively led a PQC standardization effort since 2016, leveraging a large and international research community. The effort is expected to take six or more years to vet proposals, and to select alternatives that are believed to be secure against both quantum and classical computers. Meanwhile, many point out the urgency of transition due to the threat of "record now, exploit later" in which encrypted information assets are captured and stored today by an adversary for attack later when scaled quantum computers become available.

While NIST's standardization effort aims to determine which PQC algorithms are robust enough to provide suitable alternatives for the threat of quantum computers, that effort does not address the problem of migration from today's widely deployed algorithms to future PQC alternatives across the diversity of computer systems that serve our society. Today, there are more than 4.1 billion Internet users, nearly 2 billion websites, and more than 3 trillion dollars in retail activity associated with the Internet.[5] Underlying this explosive digital transformation of the world as we know it are security and privacy technologies relying on public key cryptographic standards at many layers. The extensiveness of public key cryptography usage across the Internet means that an industry-wide migration to quantum safe cryptography standards (i.e., PQC) will be a massive undertaking, and one that is complicated by the layered complexity and heterogeneity of the worldwide compute infrastructure we operate. It will challenge, domain by domain, the fabric of our compute infrastructure and involve myriad organizations, from those who contribute widely used software and hardware components to the much larger number of operators who deploy and manage the constituent pieces of secure infrastructure globally. It is no wonder that prior history shows cryptographic migrations (e.g., 3DES to AES, MD5 to SHA1, SHA1 to SHA2, RSA to ECC) to take a decade or more before completion.

On January 31-February 1, 2019, the Computing Community Consortium (CCC) held a workshop in Washington, D.C. to discuss research challenges associated with PQC migration. Entitled, "Identifying Research Challenges in Post Quantum Cryptography Migration and Cryptographic Agility", participants came from three distinct yet related communities: *cryptographers* contributing to the NIST PQC standards effort, *applied cryptographers* with expertise in creating cryptographic solutions and implementing cryptography in real-world settings, and *industry practitioners* with expertise in deploying cryptographic standards within products and compute infrastructures. Discussion centered around two key themes: *identifying constituent challenges in PQC migration* and *imagining a new science of "cryptographic agility"*.

Key findings for **PQC migration** include:

◗ There is an important need for research to understand and quantify the implications of replacing today's public cryptography algorithms (e.g., RSA, ECDH, ECDSA, DSA) with PQC alternatives across a wide variety of implementation and deployment contexts.

◗ Given that PQC algorithms generally have greater computation, memory, storage, and communication requirements (e.g., larger key sizes, more complex algorithms, or both), research and prototyping is needed to better understand performance, security, and implementation considerations.

- Research is needed on approaches to introducing new PQC algorithms (e.g., hybrids) within live systems that must remain interoperable with other systems during the period of massive industry migration. This includes such areas as formal modeling, automated tools, and approaching transition in complex infrastructures.

Key findings for **cryptographic agility** include:

- There is a need to broaden and recast traditional notions of cryptographic agility in light of the size and complexity of global PQC migration. A new science of cryptographic agility should include an expanded set of goals, a more comprehensive set of compute domains, a broader range of agility modalities and time scales, and the full canon of security research methodologies.

- Research on cryptographic agility should include frameworks and architectures that enable agility across a wide variety of compute contexts, usable interfaces addressing various user roles, a better understanding of security and complexity tradeoffs, and other defining challenges.

- Context agility, or cryptographic frameworks that automatically select among algorithms and configuration based on the context of use, represents a long-term research vision that could shape the field.

- Cryptographic agility, independent of PQC migration, offers the benefit of making security systems more robust against algorithmic breakthroughs, revealed implementation flaws, emerging hardware accelerators, and other threats. It enables change in response to evolving security policy within an organization and support for new cryptographic features.

- In the context of PQC, it enables agility across multiple standards likely to be approved by NIST.

**Additional findings** include:

- Fundamental research is needed on policy, process, and people since these determine whether and when PQC adoption occurs at all.

- Research is needed on the frontiers of cryptography; that is, how PQC migration and cryptographic agility will apply to newer cryptography fields like secure multi-party computation, fully homomorphic encryption, blockchain, and more.

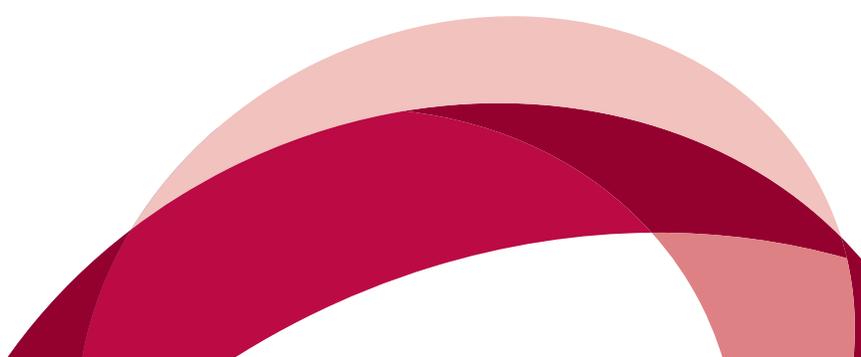

# I. Introduction: Why Post Quantum Cryptography (PQC)?

In this section, we discuss the implications of quantum computing for public key cryptography and motivations for research into the systems and issues surrounding deploying PQC in practice.

## 1.1 The Threat of Quantum Computing to Cryptography

Quantum computing, or the use of quantum mechanical phenomena to represent and manipulate information, promises to be a game-changing technology when fully realized at scale. Many problems that are now considered to be intractably complex for conventional computers, even the most powerful supercomputers, could become computable in minutes or seconds by harnessing the properties of quantum physics (e.g., entanglement, superposition) to represent information. Initial applications of quantum computing include the simulation of complex molecular systems in chemistry and material science, high-dimensional machine learning classification, and optimization problems over an extremely large space of possible solutions.

While quantum computing creates a whole new paradigm for solving complex computing problems, unfortunately it also enables a powerful new tool for attacking our existing cryptography algorithms. This makes it an important threat to Internet security as we know it today. To explain, public key (asymmetric) cryptography relies on *trapdoor mathematical functions* which allow the easy computation of a public key from a private key but make the computation of a private key from a public key (the inverse) computationally infeasible. Widely used trapdoor functions rely on the difficulty of *integer factorization* and elliptic curve variants of the *discrete logarithm* problem, both of which have no known solution for computing an inverse in polynomial time with conventional computing. In symmetric key cryptography, the security of a key shared between two parties relies on how difficult the random key is for an attacker to guess. If the value cannot be determined directly by cryptanalysis, the attacker may apply search methods to examine the space of possible keys looking for the correct value. But given a sufficiently large space of possible values, finding a key is computationally infeasible for the window of time during which the scheme is employed to protect data.

In 1994, Peter Shor showed how a quantum computer (QC), if it were to exist in a scalable form, could be used to perform integer factorization in polynomial time (polynomial in $log\ N$ on integer size $N$) using modular exponentiation by repeated squaring and a quantum Fourier transform that he designed.[1] *Shor's algorithm*, as it is now called, has been shown to generalize to also solve the discrete logarithm and elliptic curve discrete logarithm problems in polynomial time. For an attacker with a sufficiently large QC, this effectively "breaks" the security of key trapdoor functions that our widely used public key infrastructure has relied upon for years. That is, an attacker could use a QC to obtain private cryptographic keys from public keys quickly and efficiently. In 1996, Lov Kumar Grover furthermore showed that QCs could be used to solve the problem of linear search over an unsorted N-element space in $O(\sqrt{N})$ operations using a special diffusion operator that he developed.[2] For adversaries with a QC, *Grover's algorithm* implies the weakening of our symmetric key algorithms by proposing a more efficient way to search the space of possible keys in order to obtain the secret value. In 1999, Gilles Brassard *et al*. showed that QCs could be used to solve the problem of finding hash function collisions in $O(\sqrt[3]{N})$ operations using Grover's algorithm.[28]

The implications of these surprising results to public and symmetric key cryptography are well-documented. Both the NSA/CSS IAD "Commercial National Security Algorithm Suite and Quantum Computing FAQ" of January 2016 [3] and the NIST "Report on Post-Quantum Cryptography" [4] of April 2016 call out the need for new standards to replace cryptosystems based on integer factorization and discrete logarithm problems. This includes replacing widely used RSA, ECDSA, ECDH, and DSA cryptosystems with *post-quantum cryptography (PQC)* alternatives. (PQC is also known as *quantum-resistant* or *quantum-safe* cryptography.)

| Cryptographic Algorithm | Type | Purpose | Impact from large-scale quantum compuer |
|---|---|---|---|
| AES | Symmetric key | Encryption | Larger key sizes needed |
| SHA-2, SHA-3 | ----- | Hash functions | Larger output needed |
| RSA | Public key | Signatures, key establishment | No longer secure |
| ECDSA, ECDH (Elliptic Curve Cryptography) | Public key | Signatures, key exchange | No longer secure |
| DSA (Finite Field Cryptography) | Public key | Signatures, key exchange | No longer secure |

*Figure 1: Impact of QC on cryptography algorithms (Source: NIST)*

In December of 2016, NIST took a major step forward in addressing the situation by announcing a call for PQC algorithm proposals to be considered for standardization in a 6-year selection process. More than 80 proposals were submitted by the November 30, 2017 deadline, and the First PQC Standardization Conference was held on April 11-13, 2018 in Fort Lauderdale, FL. This initiative effectively organizes the cryptographic research community — including academic, industry, and government cryptography experts — into a focused effort to propose and vet the most robust alternatives suitable for new and quantum safe public key cryptography standards.

While symmetric key cryptography and secure hash functions are also impacted by the threat of quantum computing due to Grover's algorithm and the algorithm of Brassard *et al.*, increasing key sizes and output sizes, respectively, is a well-understood approach to remediate. Changes to key and hash output sizes, in practice, is highly impactful to widely deployed cryptography for data in motion and at rest and will require considerable engineering to make the transition. But our focus henceforth will be on research surrounding the newer and less understood problem of replacing our public key cryptography algorithms with quantum safe alternatives. Additionally, many point out that implementing Grover's algorithm on QCs is expected to be difficult in practice due to long-running serial computations and the need for deep circuits.[19,29]

Note the important and clarifying distinction between *Post Quantum Cryptography (PQC)* and *Quantum Key Distribution (QKD)* which are often confused. While PQC focuses on cryptographic algorithms that are resistant to QC attacks, QKD is a quantum technology for securely distributing shared cryptographic keys between two endpoints. QKD leverages the properties of quantum mechanics and is often implemented using polarized photons or entangled pairs of photons over fiber optics, although schemes also exist for free space. Technologies related to QKD include *quantum communications* which enables the exchange of qubits and entanglement between quantum computers, and *quantum networks* which looks at the use of quantum communications to connect multiple sites across larger geographic areas. Our focus here is on PQC which will be widely implemented on conventional computing systems as a safeguard against scaled QCs which are likely to exist in the future.

### 1.2 The Problem of PQC Migration: An Approaching Storm

While NIST's standardization effort is aimed squarely at the problem of determining *which* cryptographic algorithms are robust enough to provide safe alternatives for a post-QC world, there is another major challenge to consider: that of *migrating our extensive infrastructure from today's widely deployed algorithms to PQC alternatives.* We argue that far from a mere "practical consideration", this migration is in need of extensive research as a companion domain.

Today, there are more than 4.1 billion Internet users, nearly 2 billion websites, and more than 3 trillion dollars in retail activity associated with the Internet.[5] Underlying this explosive digital transformation of the world as we know it (e-commerce, e-trading, e-government, e-health, social media, smartphone apps, and more) are security and privacy technologies relying on public key cryptography standards at many layers. Cryptographic transport protocols secure end-to-end communication exchanges between Internet endpoints; public key infrastructure uses digital certificates to verify the identity of parties before private data is transferred; public key cryptography is used to seal symmetric cryptography keys in many contexts of bulk data encryption; digital signatures are used to ensure the integrity and authenticity of operating system and application software updates; cryptographic protocols are used to authenticate users and manage identities across systems and services; public key cryptography is used to securely manage public and private cloud infrastructures, including the transfer and storage of private data; hardware devices use cryptographic features to prevent data leakage and to store secrets securely; and much, much more. An illustrative list of public key cryptography applications might include Public Key Infrastructure (PKI), key management systems, authenticated web communication (TLS), secure point-to-point communication (SSH), transport security (IPSec), key agreement, identification and authentication, password-authenticated key exchange (PAKE), PGP/GPG, Secure/Multipurpose Internet Mail Extensions (S/MIME), Kerberos, Over-the-Air Rekeying (OTAR), Domain Name System Security Extensions (DNSSEC), Encrypted File Systems, Internet Key Exchange (IKE), ZRTP (a secure VoIP protocol), and more. In fact, a full description of cryptography usage domains is so large, it is beyond the scope of this report to enumerate.

The extensiveness of public key cryptography usage across the Internet (and within private networks) means that an industry-wide migration to quantum safe cryptography standards (i.e., PQC) will be a massive and global undertaking, and one that is complicated by the layered complexity and heterogeneity of the worldwide compute infrastructure we operate. It will challenge, domain by domain, the fabric of our compute infrastructure and involve myriad organizations, from those who contribute widely used software and hardware components to the much larger number of operators who deploy and manage the constituent pieces of secure infrastructure globally. It is no wonder that prior history shows cryptographic migrations (e.g., 3DES to AES, MD5 to SHA1, SHA1 to SHA2, RSA to ECC) to take a decade or more before completion.

While the realization of scaled quantum computing may seem a distant concern, there are some important reasons why the problem of PQC migration has urgency to many organizations, industries, and governments worldwide:

◗ Uncertain QC development timeline,

◗ Complex PQC migration requirements,

◗ Record now, exploit later attacks, and

◗ Relevance to NIST standards selection.

First is *risk* stemming from an uncertain quantum computing development timeline that leaves open the possibility of faster advancements than originally anticipated. NIST's 2017 PQC call for proposals notes the "noticeable progress in the development of quantum computers" as a key motivation for initiating the standards process, including "theoretical techniques for quantum error correction and fault-tolerant quantum computation, and experimental demonstrations of physical qubits and entangling operations in architectures that have the potential to scale up to larger systems." [19]

A second concern is the time and complexity of PQC migration which implies the need for considerable lead time before scaled QCs are available. NIST acknowledges this as well, stating in their PQC call for proposals that "a transition to post-quantum cryptography will not be simple as there is unlikely to be a simple 'drop-in' replacement for our current public-key cryptographic algorithms." [19] This situation is further exacerbated in embedded environments and other settings dependent on cryptographic hardware. For instance, the chief of computer security at NIST remarked that "cryptographic agility is critical for small satellite security."[35] Such systems are hard to modify and are known for long-lived deployments.

Third is concern over the possibility of "record now, exploit later" (also known as "harvest now, decrypt later") in which an adversary captures encrypted versions of IDENTIFYING RESEARCH CHALLENGES IN POST QUANTUM CRYPTOGRAPHY MIGRATION AND CRYPTOGRAPHIC AGILITYlong-lived private information assets (e.g., social security numbers, critical business information) for attack later when quantum computers become available. The threat implies the critical need for quantum safe protections within industry and government for long-lived information assets well in advance of a fully implemented threat.

Finally, there is an important need to explore PQC migration considerations during NIST's standardization process which is in full swing at the time of this writing. Research will inform NIST's evaluation of PQC proposals and help to ensure that practical standards and parameter guidelines are selected.

## 1.3 The Need for Research

The complex challenge of migrating our global compute infrastructure to new public-key cryptography standards will involve work on many levels, and we argue that *the area overall is in dire need of research*. That is, before the global industry ecosystem can deploy quantum safe solutions, there is considerable work to be done understanding migration challenges and schemes, and more rigorously addressing integration, security, performance, agility, and other challenges. The research community, known for its analytic rigor and freedom to explore, is uniquely positioned to contribute in this space. In particular, an interdisciplinary collaboration between cryptography, applied cryptography, and system security researchers is needed to understand this new and cross-cutting domain.

On January 31-February 1, 2019, the Computing Community Consortium (CCC) held a workshop in Washington, D.C. to identify the many ways in which the research community could dramatically aid the complex challenge of a global transition to PQC standards. The workshop entitled, "Identifying Research Challenges in Post Quantum Cryptography Migration and Cryptographic Agility", brought together participants from three related but distinct communities: *cryptographers* contributing to the NIST PQC standards effort, *applied cryptographers* with expertise in cryptography implementation and real-world applications, and *industry practitioners* with expertise in the deployment and usage of cryptographic standards in products and across compute infrastructures.

Discussion of research challenges focused on two overlapping domains:

1. ***Core PQC Migration Research***. Research that addresses the application of candidate algorithms to specific contexts and how migration within any given cryptographic usage domain can be realized in a secure way.

2. ***Toward a Science of Cryptographic Agility.*** Research here looks at the notion of "cryptographic agility", or the ability to migrate to new cryptographic algorithms and standards in an ongoing way.

In many ways, cryptographic agility represents the generalization of PQC migration in that it considers not just the current challenge of migrating from our current algorithms to PQC alternatives, but the longer-term need for ongoing migrations as new attacks and better algorithms motivate the need for updates in our cryptographic standards. As will be discussed in Section 3, there is an important need to develop a principled science of cryptographic agility; one that broadens and expands the scope of agility to that of developing secure frameworks that enable ongoing cryptographic advancements in a wide variety of system, protocol, and application contexts.

Figure 2 shows the relationship between PQC migration and cryptographic agility. We envision the larger space of PQC migration and cryptographic agility challenges to partially overlap. The overlap of concerns represents the space of challenges in which agility frameworks will be needed to enable cryptographic migration to PQC algorithms. At the same time, there are PQC migration challenges that are specific to the algorithms and don't involve cryptographic agility. Similarly, there are cryptographic agility challenges that are independent of PQC migration specifics.

The unique space of research opportunities is a subset of the overall challenges associated with PQC migration and cryptographic agility. This acknowledges the role of software solution providers, hardware vendors, government standards bodies, international consortiums, the open source developer community, and others who will contribute to implementing migration and agility at many levels. The challenges we are concerned with here are the parts of the overall content landscape that would benefit from the empirical rigor, analysis of fundamentals, and exploratory approaches characteristic of research.

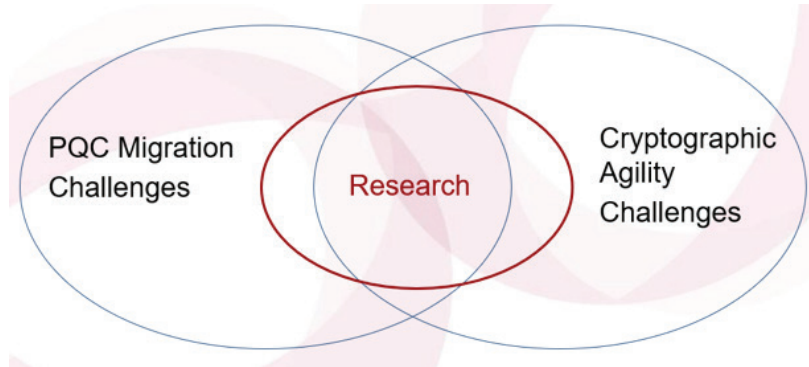

*Figure 2: The role of research in PQC migration and cryptographic agility.*

## 2. Core PQC Migration Challenges

The need for research in PQC migration begins with the need to understand the many contexts in which transition to new PQC algorithms will occur, and the problem of how migration will be implemented. This includes the problem of obsoleting deprecated algorithms, something that is surprisingly hard within a complicated world of deeply entrenched deployments and frameworks with little or no support for phased retirement.

By way of background, the five key families of PQC algorithms are shown in Figure 3. Algorithm families address three fundamental applications of public key cryptography: encryption, key encapsulation mechanisms (KEMs), and digital signatures. As NIST emphasizes in their 2016 CFP, it's important to understand that migrating our public key cryptography infrastructure to PQC alternatives will not be a simple "drop-in" replacement exercise for several reasons. First, PQC algorithms offer significantly different key sizes, ciphertext sizes, signature sizes, communication requirements, and computational requirements including both memory and compute. Second, many algorithms introduce new requirements (e.g., state management, entropy) that will demand modifications to existing frameworks. (Section 2.1 will discuss these issues in greater detail.)

Finally, NIST has indicated that there is relatively little chance that a single algorithm will be selected as the replacement. This is because different algorithms offer different trade-offs in, for example, keys size and compute requirements, and options are needed to cover a wide variety of device and usage contexts. It is also because there is naturally some uncertainty involved in quantum safe algorithm selection; the exact scope of quantum algorithms is not yet known, and the possibility of classical computing attacks is an ongoing issue for newly introduced algorithms that have not yet stood the test of time. All in all, NIST believes it is circumspect to provide several alternatives within the new PQC standard rather than forcing a one-size-fits-all solution.

Below we discuss several areas of PQC migration that could benefit from research.

### 2.1 PQC Algorithms: Charting Implications Across Domains

A core set of research challenges surrounds the need to understand and quantify the implications of replacing today's public cryptography algorithms (e.g., RSA, ECDH, ECDSA, DSA) with PQC alternatives (see Figure 3) across a wide variety of implementation and deployment contexts.

As mentioned by NIST, there are significant differences between PQC and our widely deployed public key cryptography standards. Most obvious are increases in key, ciphertext, and signature sizes which many of our current usage domains are not prepared to accommodate. But additional differences in computation and memory requirements are just as significant, impacting implementation strategies, performance, system buffering dynamics, communication patterns, and side channel vulnerabilities. Many PQC algorithms further introduce new requirements including state management (hash-based signatures), auxiliary functions (e.g., Gaussian sampling in lattices), entropy (e.g., lattice-based schemes), and nonzero decryption failure probabilities (e.g., code-based encryptions schemes). [6] How these considerations play themselves out for a wide variety of cryptography implementation and application domains represents a large open space of much-needed research.

| PQC Algorithm Family | Function/Use | Examples | Notable Attributes |
|---|---|---|---|
| **Hash-based Cryptography** | Digital signatures | XMSS, SPHINCS+ | Well-understood. Stateful schemes needed to reduce large signature sizes. |
| **Lattice-based Cryptography** | KEM/Encryption, Digital signatures | FrodoKEM, NewHope, NTRU, FALCON, qTESLA | Short ciphertext and keys, good performance, sometimes complex. Short signatures. |
| **Code-based Cryptography** | KEM/Encryption, | BIKE, Classic McEliece, HQC, NTS-KEM, RQC | High confidence. Fast encryption but larger public keys. |
| **Multivariate Cryptography** | Digital signatures | EMSS, LUOV, MQDSS, Rainbow | Large key sizes (~1 MB / ~11 KB). Schemes need more analysis |
| **Supersingular Elliptic Curve Isogeny Cryptography** | KEM/Encryption | SIKE | Very small key sizes (less than 500 B), slower performance, relatively new. |

*Figure 3: Families of PQC algorithms and key attributes [10]*

A non-exhaustive list of traditional cryptography usage domains for exploration includes:

◗ secure communication protocols (e.g., TLS, SSH, IPSec),

◗ digital signature schemes,

◗ public key infrastructure (PKI),

◗ authentication protocols,

◗ identity and access management systems, and

◗ key management systems.

The space of cryptography usage domains can also be looked in a platform- and/or application-centric way. For example, research might consider how deploying PQC algorithms will impact:

◗ web-based computing,

◗ mobile computing,

◗ Internet of things (IoT) and edge computing,

◗ public, private, and hybrid clouds,

◗ virtual private networks, and

◗ trusted computing architectures.

In general, workshop participants pointed out the need for research to be highly experimental. Software and hardware test environments are needed to prototype and quantify experimentally what happens when PQC algorithms are deployed within a broad range of existing cryptography domains.

## *Performance Considerations*

Given that PQC algorithms generally have greater computation, memory, storage, and communication requirements (e.g., larger key sizes, more complex algorithms, or both), research is needed to better understand and quantify performance considerations in a wide range of deployment contexts. Broadly, performance is a key industry concern, and an important set of challenges to be solved before PQC can be adopted in practice.

Consider networking, for example. Larger key sizes imply changes to packetization and latency patterns within secure communication protocols like TLS. This in turn impacts a whole spectrum of network-related devices that have been optimized for our current cryptographic protocols in the interest of performance and scale – from network routers and switches to gateway devices, network appliances (e.g., firewalls, intrusion detection systems, WAN accelerators), and content distribution schemes. How will packetization considerations impact network function virtualization in 5G cellular networks? What are the implications of new PQC communication patterns for end user devices like smart phones or end user applications like web browsers?

Research could play a much-needed role in developing *performance optimization* approaches for specific PQC algorithms. Where are the key bottlenecks for a given algorithm, and what frameworks might be developed to address them? Examples include new ways to exercise parallelism, new data structures that improve memory access performance, or mathematical techniques that reorganize computation to better utilize an underlying compute architecture. Included in this research challenge is the need to understand performance in IoT device contexts where compute, memory, and battery constraints become first order considerations.

Techniques for making PQC more performant can also be applied to the challenge of *hardware acceleration*. FPGA-based research could be used to explore accelerator designs for key memory and computation bottlenecks in various families of PQC algorithms. As seen over time with AES and SHA-2, hardware primitives could be designed that lead to new instruction proposals for widely used computer and communication architectures.

### *Security Considerations*

Changes to the characteristics and requirements of our public key cryptography algorithms are more than just a matter of performance. They create new security issues in a variety of ways.

In contrast to well-understood RSA and ECC algorithms, less-understood PQC candidates have a different set of trade-offs in configurable parameters such as key size, ciphertext size, and computation time. Furthermore, specific algorithms add new "knobs"; for instance, dimensions in lattice schemes or code length and dimensions in code-based schemes (e.g., Classic McEliece). A key challenge exists in understanding the tradeoffs between security and algorithm requirements for a wide variety of usage domains. These tradeoffs are unlikely to be addressed fully by NIST who cannot consider all contexts of PQC algorithm usage. While NIST will standardize schemes with specific parameter settings, guidelines on selecting algorithms among multiple options, and on security levels for specific usage contexts will be needed. Research may also lead to parameter adjustments over time.

Another area of much-needed research is in the *cryptanalysis* of PQC algorithms across a wide variety of protocols contexts. In cryptanalysis, researchers look for weaknesses in a cryptographic scheme under various assumptions and adversary models, and ultimately what it takes for a given cryptographic scheme to be broken. Research is needed on threat models in the context of specific PQC algorithms. Cryptanalysis could include both analytic components (e.g., careful investigation of underlying hardness assumptions and how they are mapped to real-world implementations) and practical components (e.g., strategies for attacking a scheme using statistical means or examining the security impact of a weak entropy source).

An important cryptanalysis challenge is that of *side channel vulnerabilities* or information leaks surrounding specific hardware architectures. In general, individual PQC algorithms will introduce new patterns of memory usage, timing, communications, cache behavior, failure modes, and more. How these patterns can be used by an adversary for timing attacks, memory-based attacks, differential fault analysis, speculative execution attacks, and other types of side channel attacks is an open question. Work is needed that examines such possibilities (and mitigations) for a broad spectrum of hardware platforms, from multi-socket servers to widely used end user devices to rapidly proliferating IoT devices.[1]

### *Implementation Considerations*

The implementation of cryptography, whether in software or hardware, is notoriously more difficult than it appears. In part, this reflects the complexity of mathematical algorithms, which are a common source of errors. More fundamentally, however, it reflects the difficulty in translating mathematical algorithms to platform-specific architectures and device contexts. For example, the details of data representation and layout, and its interactions with a system's memory hierarchy and operating system buffering mechanisms, can introduce vulnerabilities that are not apparent within cryptographic algorithm design.

Given these complexities, there is an important need for research exploring the *implementation of PQC algorithms*

---

[1] The IoT space, in particular, suffers from the pernicious synergy of great cost-sensitivity, attachment to physical devices, diffcult-to-manage embedded interfaces, and insufficient security incentives for vendors.

across a broad range of devices, computer architectures, system software stacks, and programming languages.

A particularly challenging context of implementation is that of embedded systems. Devices in this domain are constrained in memory size, compute resources, and power availability since battery lifetimes are finite. PQC implementations are needed to understand how specific algorithms can navigate such constraints, and how hardware-software boundaries should be defined to help. Since IoT devices are physically exposed to adversarial tampering, how can implementations help to guard against side channel attacks in various ways? Which PQC approaches, and which parameter choices, are well- or poorly-matched with devices in this domain?

Note that existing reference code associated with NIST PQC submissions is often not ready for real-world use within industry contexts and, while optimized implementations were additionally submitted, there is an important need for further research on optimization and performance. An important outcome of research could be a common set of robustly implemented, optimized software libraries (e.g., within Open Quantum Safe [32]) to support experimentation and performance characterization on specific platforms and using specific programming language runtimes.

## 2.2 Migration Frameworks: How will we get there?

Migration from today's widely used public key cryptography algorithms to PQC replacements is not merely about the algorithms. Another much-needed area of research surrounds the *approaches used to introduce new algorithms within systems that must operate continuously and remain interoperable* with other systems that may be ahead or behind on the migration curve.

One widely discussed approach to introducing migration is that of *hybrid* schemes. In this approach, two cryptographic algorithms are applied, one from our current canon of standards (e.g., RSA or ECC) and one from the newer array of PQC alternatives (e.g., lattices). Hybrids provide a way to introduce quantum safety to address "record now, exploit later" while still relying on well-understood resistance to classical attacks. This is especially important during initial periods of PQC deployment since confidence in the robustness of newly introduced algorithms and implementations takes time to build. An example can be seen in X.509v3 digital certificates which support an option enabling embedded extensions. [16] The option can be used to embed a PQC public key and signature within a digital certificate using conventional standards. A nice feature of "hybrid modes" of cryptography usage is that an organization can retain certification (e.g., NIST FIPS 140) during the period of transition to newer candidate standards. A drawback, of course, is that computation, memory, communication, and other requirements are significantly increased. It was noted by workshop participants that migration to PQC alternatives may, in fact, involve two migrations: one to hybrid schemes and a subsequent migration from hybrid schemes to standalone PQC algorithms. This likely future further underscores the need for cryptographic agility which will be discussed at length in Section 3.

Some research on hybrid schemes has been carried out, specifically in the context of key exchange protocols.[30,31] Hybrid schemes are related to cryptographic research on *combiners*. Hash combiners construct a new hash function from two component hash functions and exhibit robust security if at least one of them is secure. [17] Encryption combiners, used with identity-based encryption schemes, take public keys from component encryption schemes and create a combined public key. [15] Work on combiners, to our knowledge, has not been applied more specifically to PQC algorithms and represents an important area of future work. For instance, a better understanding of combiners and key derivation functions (KDFs) is needed for deploying hybrid schemes.

Another approach to migration is cipher suite negotiation as seen in IETF protocols like TLS. [18] A list of supported cipher suites is presented by interacting parties during the protocol's initial handshake phase in order to select the most robust option that both support. The negotiation may include cipher suite version numbers and additional information on key sizes and parameter settings. [14] Using this framework, new PQC algorithms could be added as cipher suite options and deprecated algorithms removed. As will be mentioned later in section 3.1, downgrade attacks are a concern in this approach.

In general, while practical schemes exist and are even being discussed by some standards bodies, there is a need for research to examine migration frameworks more creatively and more rigorously. For each domain and platform type, what new migration approaches could be developed to support the transition to new PQC algorithms without loss of interoperability and functionality during the transition period? What is the attack surface and risk profile associated with each approach?

A key problem for any migration scheme is that of legacy systems. How can systems that are not designed to be migrated make the transition to PQC algorithms? Examples include legacy IoT devices that are no longer supported by a manufacturer or legacy system software that is no longer actively being developed or supported. Since public key cryptography involves the secure interoperability of systems, how should a migrated system interact with a legacy system in a quantum safe manner? Are there frameworks that can be applied transparently to protocols or systems that lack inherent migration mechanisms? What options exist for adding PQC to systems that cannot be field-upgraded?

Many workshop participants also noted the problem of algorithm *deprecation*. Prior cryptographic migration efforts have shown that eliminating deprecated algorithms (e.g., RC4, MD5, SHA1, DES and TripleDES) from active use is harder than it appears. Legacy systems, legacy versions of cryptographic libraries, lack of security oversight within an organization, and many other factors contribute to this. The research question to be addressed is how algorithm deprecation can be "designed in" to a migration scheme to ensure deprecated algorithms don't operate in perpetuity.

Finally, workshop participants pointed out the need for research to address the problem of *when* migration to PQC should occur. While timeline discussions often look to the current state of QC prototype development, this fails to account for the complexities of migration and the chain of dependencies between research, standards bodies, hardware platform providers, software solution providers, open source libraries, system software stacks, industry certification frameworks, and more. Research on risk management could be part of investigations in this arena.

## *Formal Modeling*

An area of much-needed research is *formal modeling of cryptographic migration*. While industry can build migration schemes, whether hybrid or combiner-based or negotiation-based or something else, it remains an open question whether the presumed security has the robustness that is expected. Formal methods can help to address this question for a given scheme in a more fundamental manner and leveraging analytic techniques that researchers are uniquely positioned to apply.

For example, consider the need for formal modeling of hybrid and combiner schemes. While the intuition is that they are no weaker than the strongest underlying component algorithm, can we be sure they don't introduce vulnerabilities compared to a non-hybrid use of the strongest underlying algorithm? What are the attack surfaces associated with specific hybrid instantiations for key encapsulation mechanisms, encryption schemes, and digital signatures? What do formal models tell us about the level of security for common misuses or flawed implementations? How are adversaries to be modeled under a variety of assumptions?

Formal modeling is also needed for examining the security of inserting migration frameworks into common cryptographic protocols. An important example is the widely used TLS protocol which could be carefully divided into classical and quantum resistant components for analysis. Other protocols include key exchange (Diffie-Helman), key management (KMIP), public key infrastructure (PKI), secure communication (IPSec, SSH), secure web or mail applications (HTTPS, S/MIME), signature applications (MAC, PGP, CMS), identity management (SAML, OpenID), virtual private networks (VPNs), and more.

## *Automated Tools*

Workshop participants pointed out that the extensiveness of our cryptographic infrastructure makes migration nearly unachievable without automated tools. As such, there is an important need for research to address this issue in a variety of spheres.

One important challenge is identifying where public key cryptography is being used in an organization's complex IT infrastructure and which algorithms and versions are deployed. While this may seem simple, the challenge speaks to how complex and multi-layered our uses of public key cryptography are. Large organizations may leverage hundreds of software packages and develop many more. Application binaries may have chains of dependencies that extend to third party application providers who in turn use third party component libraries, cryptographic modules, operating system APIs, vender-specific device driver libraries, and so on. Research could help to develop active and passive approaches to scanning an infrastructure and provide an analysis of legacy cryptography usage based on network traffic, open ports, end user devices, system binaries, source code repositories, and more.

Automation is also needed for analyzing PQC migration mechanisms inserted into protocols, specifications, and source code. Application source code, for example, is often too large for manual scanning and too complex for programmer analysis. Automated tools could trace dependencies, identify runtime control flow, probe for common vulnerabilities, and verify the security of new PQC libraries and migration mechanisms. Because of the strong association with formal modeling, once again researchers are uniquely suited to advance the state-of-the-art in this sphere.

Automated testing tools are sorely needed to test PQC migration mechanisms, and to explore cryptographic failure modes, whether PQC algorithm specific or migration framework-based. Forward-looking automation research could develop frameworks for synthesizing migration and validation code, inserting test cases into the developer toolchain, modifying binary images for legacy software, and so on.

*Complex Infrastructures*

Very little research has been done in understanding PQC migration challenges in complex compute infrastructures like private data centers, public cloud, hybrid and federated architectures, edge computing, smart home or building environments, and more. Such infrastructures not only exhibit architectural complexity, they deepen the layering of our system software stacks and add heterogeneity.

Workshop participants pointed to the need for research on the software stack implications of PQC migration. While migration may seem as simple as changing a library at a single layer in the stack, in fact, there are often implicit dependencies that introduce complexities. For example, digital certificates may be parsed at the application layer, cryptographic keys may be managed by an infrastructure management agent, or network security mechanisms may be tuned to particular packet sequences for a given cryptographic protocol. Research is needed to develop mechanisms that allow a better understanding of such cross-layer dependencies.

Research is also needed on infrastructure-level abstractions and frameworks for addressing PQC migration. Given a complex web of migration domains, which are the most important and what might a priority ordering look like? Which key dependencies would result in the greatest impact if they were migrated first? In general, how might we model the migration of an entire infrastructure? How should migration auditing work, and how might we construct evidence of end-to-end quantum safety? What tools might identify and address infrastructure-level vulnerabilities and failures?

On a macro level, research will be needed in the future on how to measure the state of PQC migration across a geographic region of the Internet, or even the global Internet itself. Methodology and tools do not currently exist.

### 2.3 Case Study: Authentication Using PQC

Workshop participants pointed out that one of the biggest challenges in migrating to PQC alternatives is that of authentication. Today's digital signatures standards, as defined by NIST [20], include DSA, RSA, and ECDSA algorithms, which, for a security level of 112 bits, offer key sizes of 2048, 2048, or 224 bits respectively [33]. Algorithms are generally fast and efficient, particularly ECDSA which reduces both key size and computation requirements.

Finding a quantum safe alternative for these widely deployed signature algorithms illustrates some of the complexities that could benefit from research on PQC migration. Multivariate signatures schemes (e.g., Rainbow [23]) offer small signature sizes, but large public and private keys (e.g., 500-700KB) and significantly more computation for key generation and signature operations.

Lattice-based schemes (e.g., CRYSTALS-Dilithium [24]) offer key sizes that are roughly equivalent to DSA and RSA, but still require significantly more computation for signatures.

Hash-based signatures, another quantum safe alternative, were proposed in the 1970s [25] and offer well-understood security under comparatively few assumptions. However, they introduce the new complexity of statefulness since one-time signatures require that a secret key not be used twice. To solve the problem of large public key sizes, hash trees (i.e., Merkle Trees [26]) can be used to generate a large number of pre-computed keys of small size for signature use. Now, however, users of the scheme need to securely store and manage the private keys associated with such trees, a consideration that makes migration from DSA or ECDSA challenging. While stateless hash-based PQC alternatives are being considered (e.g., SPHINCS+ [27]), keys sizes are much larger. [6]

The implications of these ins and outs to a broad range of authentication applications remains an open question to be explored by researchers. On the one hand, it's unclear how larger storage, memory, and communications requirements will impact various domains if stateless algorithms are chosen. (Consider, for example, IoT devices.) On the other hand, stateful schemes change the requirements of cryptography usage dramatically, necessitating new implementation schemes to deal with one-time key pairs and to avoid attacks directed at key reuse. Workshop participants pointed out that stateful schemes (e.g., XMSS [28]), despite some appealing properties, would appear to be a poor match for many distributed computing scenarios, usage contexts that emphasize resilience, and hardware-based security schemes. Even use with PKI certificate schemes would seem difficult.

## 3. A New Science of Cryptographic Agility

The challenge of migrating our public key cryptography to quantum safe alternatives naturally raises broader questions about the very frameworks we use to deploy and configure cryptography across the global Internet. How amenable are our systems to changes in cryptography over time as algorithms, implementations, and standards continue to evolve?

Many have pointed out the widespread problem of security solutions that fail to comprehend and provision for the full *lifecycle* of cryptographic algorithms. A given cryptography standard may be revised over time as advances in technology weaken the strength of once accepted key sizes and other configuration parameters. New algorithms may be introduced in response to vulnerabilities or as more efficient alternatives become available. Standards may become deprecated and need to be eliminated in a phased or sometimes immediate way. [21] Algorithm elimination is a particularly notable failure in the industry as long-deprecated standards (e.g., RC4, MD5, DES) continue to be in use.

In the context of PQC, NIST's Lidong Chen points out in a 2017 *IEEE Security and Privacy* article [6] that new QC algorithms could be discovered in the future that lead to attacks on PQC algorithms thought to be quantum resistant. In fact, the performance characteristics and algorithmic techniques of tomorrow's QCs are open to debate as the limits of quantum computing are not yet known. This creates an unavoidable uncertainty in the longevity of upcoming PQC standards. Adi Shamir adds that NIST's PQC standards process should simultaneously comprehend the need for near-term solutions that are production ready, schemes that offer improvements but require further analysis and vetting, and longer-term research on new families of algorithms that could offer superior robustness and provability.[7]

Even NIST's current work on PQC standardization is expected to lead to several alternatives for use with encryption, key encapsulation, and digital signatures. [19] As mentioned in section 2, this is because different algorithms offer different trade-offs in key and ciphertext sizes, compute requirements, communications overhead, and so on, and options are needed to cover a wide variety of device and usage contexts. Over time, additional algorithms, parameter recommendations, and implementation options may become available for a given set of technical requirements. Change will be needed in many dimensions.

Given these considerations, we believe the industry challenge of *cryptographic migration* should be understood as a broader challenge of *cryptographic agility*; the question is not how should we transition to a new set of standards, but how should we transition to architectures that offer agility for ongoing cryptographic migrations over time? Cryptographic agility addresses the important problem of future-proofing our global cryptographic infrastructure in a flexible and robust manner.

Surprisingly, traditional notions of "cryptographic agility" within the research community invoke a fairly narrow set of concerns surrounding algorithm implementation; that component algorithms within an implementation should be replaceable, something readily addressable through engineering best practices. While an important point in systems engineering, we believe this limited view fails to comprehend the larger picture of agility challenges facing our global cryptographic infrastructure. In fact, this minor *implementation* challenge should be recast as a major *design* challenge; and one that is a key opportunity for the industry as it moves to replace deeply entrenched standards with newer PQC alternatives.

A key contribution of the CCC workshop, then, is to call out the need for *broadening and recasting the scope of cryptographic agility* in light of the size and complexity of global PQC migration challenge. Could there be a principled science of cryptographic agility that more rigorously considers a broad spectrum of frameworks, a robust analysis of correctness and security, a deeper understanding of attack surfaces, and an exploration of domain-specific (e.g., protocol, application, system) issues? We outline what we see to be the research issues below.

### 3.1 Definitions, Goals, and Scope

What should an expanded and more rigorous science of *cryptographic agility (CA)* consist of? We believe there is an important need for the research community to redefine the scope of CA in several dimensions, which we describe in this section. The challenge here is to adjust our traditional notions of CA to better comprehend the scale and complexity of cryptography usage in the wild across our global Internet.

As a first step, workshop participants pointed out the need to define a broad set of goals to guide advancement of the field. While the list is an open research question to be addressed, some illustrative examples include:

◗ *Effectiveness*. CA must be demonstrably effective in facilitating cryptographic migration.

◗ *Measurability*. The level of CA can be clearly assessed for any given algorithm or implementation.

◗ *Interpretability*. CA requirements and techniques can be applied across a range of cryptography contexts.

◗ *Enforceability*. CA techniques are well-specified and can be mandated for specific cryptography contexts.

◗ *Security*. CA approaches are secure against attacks of various types.

◗ *Performance*. Overheads caused by CA are well-understood and within acceptable limits.

A second step would be an expanded notion of the compute domains over which CA frameworks will be applied. The range of contexts should be comprehensive in addressing the universe of compute domains within which we deploy and use cryptography in practice. One way to look at such contexts is as a hierarchy of expanding scope; how might agility be defined over the following units of cryptography usage:

◗ *An algorithm* (e.g., key encapsulation mechanisms),

◗ *A unit of program code* (e.g., an authentication function),

◗ *A protocol* (e.g., TLS),

◗ *An application* (e.g., an email or web server),

◗ *A service* (e.g., an online banking portal),

◗ *A system* (e.g., an operating system or IoT device),

◗ *A distributed compute infrastructure* (e.g., an enterprise),

◗ *A cloud hosting service* (e.g., public and/or private cloud), or

◗ *A complex vertical domain* (e.g., a smart building)?

A third step in broadening the scope of CA is to comprehend modalities of "agility" beyond simple algorithmic agility. Workshop participants suggested the following additions:

| Modality | Scope |
|---|---|
| *Implementation Agility* | Application interfaces and policy configuration frameworks facilitate migration across implementations. |
| *Compliance Agility* | Cryptographic infrastructure can be reconfigured to address compliance requirements for varying international regulations and frameworks, or to minimize a trusted computing base (TCB) |
| *Security Strength Agility* | Many PQC algorithms require different implementations for different security strengths. Algorithms that dynamically ascale security strength based on configuration provide better agility |
| *Migration Agility* | The ability to move automatically from one scheme to another - including conversion. Requires better use of cryptographic metadata at the level of application data. |
| *Retirement Agility* | The ability to enforce the retirement of obsoleted or insecure cryptographic algorithms. |
| *Composability Agility* | The ability to combine cryptographic building blocks in a secure way. |
| *Platform agility* | The ability to use assured cryptographic algorithms across different platform types. |
| *Context Agility* | The cryptographic algorithm and strength policy should ideally have the flexibility to be derived automatically from system attributes such as data classification and data location. |

*Figure 4: Possible modalities of an expanded notion of cryptographic agility.*

A fourth step is to consider a broader range of time scales over which CA approaches will be applied. *Coarse-grained* approaches might consider the problem of migrating a hardware device or cloud infrastructure from one cryptographic standard to another in a phased way while ensuring continuous operation throughout. *Fine-grained* approaches, on the other hand, might address the problem of selecting a cryptographic standard instantaneously during a configuration sequence or as part of a session-based negotiation between communication endpoints. The latter might include CA schemes that let us switch cryptographic algorithms in real-time or almost real-time at any point of our operations. Broadly, CA research should address the need for agility across a range of time horizons and include time interval explicitly as a first order design parameter.

Finally, a new and principled science of CA should mean expanding the range of research methodologies to include the full arsenal of computer science research. In other words, beyond engineering best practices looking at reconfigurability, research on CA should include *formal analysis* approaches that explore more fundamental properties of a given scheme, *architectural* approaches that consider hardware-software co-design alternatives, *performance* approaches that quantify the impact of CA for a given platform or architecture, *systems* approaches that examine the layering of CA mechanisms within a cloud or networking software stack, and more.

## *Frameworks*

At the core of CA research are the frameworks and architectures that enable agility across a wide variety of computing contexts. What might such frameworks look like and what prior work might serve to bootstrap research in this domain? Overall, there is a notable scarcity of research in this area and work is needed to fill out the picture of potential approaches and to develop insights on what is possible.

A review of existing work on agility frameworks might start with widely used cryptography libraries like OpenSSL, Bouncy Castle, and Oracle JCE, all of which offer industry-ready implementations of cryptography standards and software-based frameworks for selecting among them. It is reasonable to speculate that support for PQC algorithms will be added over time as NIST standards become available. However, it is important not to confuse the availability of cryptographic libraries with *cryptographic agility*; the problem lies in the manual effort required for developers to integrate new library options into existing software, and to deal with considerations as described in section 2.1 above.

*Hybrid* cryptography schemes were mentioned in section 2.2 as a way to migrate from one standard to another. Essentially, two cryptographic algorithms are applied, one from our current canon of standards (e.g., RSA or ECC) and one from the newer array of PQC alternatives (e.g., lattices) as a way to introduce quantum safety while still relying on well-understood resistance to classical attacks. X.509v3 digital certificates were cited as a canonical example. [16] Agility research using hybrids (and cryptographic *combiners*) might consider their role as building blocks within a larger framework that addresses ongoing transitions over time. Perhaps the approach could be applied to agility contexts with other time scales (e.g., session-based, short-term vulnerability response), and variants of the scheme could be developed to address performance overheads and resource requirements, two key drawbacks cited earlier.

The existing practice of *cipher suite negotiation*, as seen in IETF protocols like TLS [18], represents an important and widely used agility approach within the communication protocol domain, enough so to warrant an IETF RFC on the subject [14]. As described in section 2.2, a list of supported cipher suites is presented by interacting parties during the protocol's initial handshake phase in order to select the most robust option that both support. The scheme addresses the need for session-based agility, and supports both the introduction and elimination of cryptographic algorithms. A principled science of cryptographic agility might provide a more robust analysis of the paradigm, including an analysis of attack surfaces (e.g., downgrade attacks). Could the approach be applicable beyond communication protocols? What happens when the approach is considered for a broader range of infrastructure and agility contexts?

We argue that the frameworks mentioned here, while important, represent a relatively small sampling of possible approaches and there is considerable opportunity for research to expand the range of alternatives. As described in section 3.1, CA considerations apply to a broad range of compute domains and modalities, encompassing a diverse set of requirements that is beyond what hybrids or cipher suite negotiation can deliver as approaches. For example, at the algorithmic level, how might agility be built into key exchange or message authentication algorithms in provably secure ways? At the infrastructure level, what is the architecture supporting cryptographic reconfiguration in a secure and auditable manner?

### Interfaces

A companion research issue surrounding that of CA frameworks is the challenge of *interfaces*. What should the interface of a CA scheme, as presented to a programmer, a security administrator, a device owner, a service provider, a systems integrator, etc. look like? What level should cryptographic configurability be handled at, and what security mechanisms should be designed-in to protect the scheme?

Workshop participants pointed out the need to address both the syntax and semantics of an interface within the context of an agility scheme and its users. The interface should be amenable to formal analysis which can be used to assert provable security properties. It should also be amenable to testing and validation.

An open question is that of *abstractions*. Research is needed to develop new user-facing abstractions for CA, and for studying the tradeoffs of different levels of abstraction for a given domain. How much flexibility should be offered and at what expense? Should multiple interfaces be offered for different user types? How should defaults be handled, something sorely needed for naïve users but known to cause problems over time as values become outdated or fail to align with a given context of use?

We note that little work has been done to explore the interface design space for cryptographic agility. Research is needed to develop design alternatives and insights on user behavior for a variety of cryptographic usage contexts. Meaningful paradigms should address a range of users, from developers to security administrators to device manufacturers to auditors and more.

## Security and Complexity Tradeoffs

An important concern and companion research issue in CA schemes is that of *attack surfaces*. It was mentioned in section 2.2 and 3.1.1 that IETF RFC 7696 ("Guidelines for Cryptographic Algorithm Agility and Selecting Mandatory-to-Implement Algorithms") [14] offers a protocol-centric view of CA, defining a successful realization to be "when a protocol can easily migrate from one algorithm suite to another more desirable one, over time". The RFC describes numerous requirements for CA schemes, including standardized identifiers and version numbers for algorithm suites, specifications for which algorithms must be implemented, mandatory key sizes and parameter settings, integrity protections for algorithm negotiation, and more.

While CA as described may seem a panacea of sorts, a key concern mentioned in the RFC is that of added complexity and the corresponding opportunity it creates for attack. For instance, a well-known attack in negotiation-based protocol exchanges between endpoints is that of the *downgrade attack* in which a man-in-the-middle adversary causes an endpoint to choose a less secure cryptographic suite option or even switch to an unprotected mode of communication by tampering with options during algorithm negotiation. Other weaknesses are created when less secure algorithm suites or less tested implementations are offered as options to enable interoperability.

There is an important need for research on attack surfaces associated with CA schemes and interfaces. Workshop participants raised the question of whether "agility builds fragility" in cryptographic systems? Paul Kocher, at a 2016 NAS Cyber Resilience Workshop on "Cryptographic Agility and Interoperability" underscored this point by stating, "Agility mechanisms introduce complexity, which leads to unknown consequences." [22]

What are the adversarial implications of introducing CA mechanisms, and how might these considerations guide the design of particular mechanisms? In a related point, workshop participants pointed out the need to prevent CA schemes from enabling bad algorithms and malicious implementations for a given usage domain.

## Other Defining Challenges

Workshop participants mentioned a number of additional research challenges.

First is that of identifying the right areas within a cryptographic solution to insert agility. The science of CA should predict where future security problems could occur within a cryptographic protocol or system, and then design the system in a way that leverages agility as a solution "hook". For instance, a potential weakness or vulnerability in a software solution could be componentized within the architecture and placed within a broader CA framework to allow the introduction of alternative components in the future. This approach to refactoring a system should, furthermore, encompass all aspects of cryptography, including configuration, management, logging, entropy sources, key management, key derivation, key distribution, and more.

Another key issue is that of testing and validation. Workshop participants pointed out that CA schemes, regardless of their scope (e.g., an algorithm, a cloud infrastructure) should come with companion testing and validation designs. How do we safely test systems with a new cryptographic standard or library implementation that deprecates support for an older cryptosystem? What is the best way to build testing frameworks that allow developers to future-proof their cryptography usage should that library or device be required to migrate? Research in this area should include automated tools for checking and enforcing CA design goals.

Yet another issue is how to address CA in legacy devices and cryptosystems that are difficult or impossible to reconfigure. A showcase example is IoT devices that are no longer supported by their manufacturer, and compute platforms with algorithm-specific security features and instruction sets. In the software domain, the problem can be seen in legacy applications or system software distributions that lack CA hooks and/or are no longer supported. Could agility schemes be built that "wrap" legacy algorithms in updated alternatives or somehow insert agility into an existing architecture?

## 3.2 Context Agility: A Research Vision

One way to explore research challenges in CA is to consider future visions that identify non-incremental leaps and longer-term goals for the field. We believe that *context agility* could provide one such vision.

The notion of context agility, as discussed by workshop participants, considers cryptographic agility frameworks that automatically select among algorithms and configurations based on their context of use. For example, a context-aware CA framework may use data classification (personally identifiable information vs. public data) or data location (within a protected campus network vs. public location) to drive algorithm and parameter choices, managing tradeoffs in specific ways.

Context agility schemes might consider CA from a wide variety of perspectives. A context-aware scheme could choose the right set of algorithms and parameters for a particular regulatory environment, or based on the availability of country-specific algorithms. Context agility could be aware of the underlying device platform and make decisions that better comprehend available compute and communication resources and the state of the system. Context agility could be used at a variety of time scales, from session-based cipher suite selection to infrastructure migrations over larger time horizons.

An important opportunity is context-aware CA schemes that could address notorious problems surrounding the users of cryptography. One of these is the failure of users to configure cryptographic systems with the right algorithms and parameters, something they often don't understand or ignore due to complexity or infrastructure scaling issues. Automatic selection of defaults (algorithms, security levels, key sizes, etc.) is a simple but important example. Another problem is that of cryptography evolution over time – introducing new algorithms, eliminating deprecated algorithms, altering parameters over time, managing new cryptography requirements, and so on.

Context agile schemes could also move the industry beyond CA frameworks that merely *select among algorithms* to a world in which agility frameworks help to d*eploy new cryptographic technologies and features*. Given the right level of abstraction, a context-aware scheme could be configured to recognize system requirements and select the right technology for a secure result. This might be particularly useful in situations where security requirements are unanticipated or introduce new requirements in an ongoing or unexpected way.

Context-aware CA frameworks could also provide an approach to recommending or auditing cryptography configurations. For example, a framework might recommend a regional (location-specific) configuration for service providers, assist IT departments with the configuration of a newly installed resource, advise a naïve user on cryptographic configuration for their home device, or provide guidance to an IT team on changes to their infrastructure relative to new regional requirements.

We believe that research is needed to explore the possibilities of self-configuring cryptography, a vision that could drive a new generation of cryptographic protections across a variety of infrastructure types and time scales.

## 4. Additional Research Directions

In this section, we discuss additional research challenges.

### 4.1 Policy, Process, People

While the technical challenges surrounding PQC migration and cryptographic agility are considerable, many workshop participants additionally pointed out that research is desperately needed to better understand people, process, and policy aspects of the problem. While technical solutions are essential, these determine whether or not adoption occurs at all and when.

An important challenge is how to create incentives for software vendors and developers to build cryptographic agility into their solution architectures. For development teams, time and effort spent on cryptography is like a tax — an overhead to be minimized or avoided because it lacks payoff. That is, unlike a new product or adding features to an existing product, cryptographic agility as a practice doesn't translate to revenue for a product team and therefore doesn't make economic sense in terms of time and effort investment.

The role of government, customers, and the Internet marketplace more generally in creating the right incentives is an important gap that could benefit from economic and policy research. For government, could cryptographic agility be incentivized through procurement policy or by early adoption programs within the context of myriad service environments? From a standards point of view, could existing FIPS-140 certification processes be modified to include cryptographic agility as a requirement for maintaining the certification of a cryptographic solution? What can government do to create the right incentives?

Workshop participants pointed out that adoption incentives within the industry often boil down to customer pressures; company priorities are often adjusted to directly reflect customer needs and requirements. As such, what possible frameworks could be developed to enable customer pressure on industry solution providers to include cryptographic agility in their products and services? More broadly and from a solution provider point of view, what do companies gain in the marketplace by having robust security and privacy? What are the right economic and policy frameworks to increase such gains in the marketplace?

Note that adoption of cryptographic agility in the manner being discussed here assumes that we have clear technical definitions and associated metrics for evaluating an implementation. Unfortunately, we currently have neither. An important research challenge is thus developing a clearer understanding of these notions and translating them to a more specific set of requirements that can be acted upon by the industry.

Finally, how to incentivize PQC adoption (when the time is right) is a key open challenge and one that is complicated by the fact that the timetable will vary for different industries and entities. Must it be fear that motivates organizations to take action, much like "Y2K" motivated the industry during that era of history? Are there education components to the problem, or would they matter little without more tangible threat and compliance incentives? Note here the "record now, exploit later" threat in which encrypted data can be captured and stored by an adversary now for attack later when scaled QCs become available.

What kind of services might be a good match for early deployment of PQC? Are there services that could be offered for free as a way of bootstrapping industry adoption? For example, workshop participants imagined a fictitious "Let's Encrypt" campaign offering free hybrid certificates for use by Internet browser providers.

## 4.2 Frontiers of Cryptography

In addition to conventional cryptography use cases, workshop participants also discussed research challenges associated with emerging areas of cryptography. Below is a list of key areas.

◗ *Secure Multi-Party Computation (MPC)*. MPC enables multiple parties to jointly compute the output of a function over private data sets in a way that maintains the secrecy of input data and ensures correctness even if adversarial parties collude and attempt to attack the protocol in various ways. MPC is also referred to as privacy-preserving computation.

◗ *Identity-based Encryption / Attribute-based Encryption (IBE/ABE)*. In IBE, a unique set of information about the identity of a user (e.g., email address) is used in place of a conventional public key. The recipient of an encrypted message then uses a trusted central authority to obtain a decryption key. ABE schemes constrain the ability of user to decrypt a message using attributes (e.g., organizational role, service tier). In other words, a message is encrypted using a set of attributes and can be decrypted only by a user who holds the private key matching the attribute formula. The scheme also makes use of a trusted party. [13]

◗ *Fully Homomorphic Encryption (FHE)*. Homomorphic encryption protects the privacy of data by enabling computation directly on ciphertexts allowing, for example, private data to be outsourced for processing in cloud computing or third party service contexts. While partially homomorphic encryption techniques constrain the scope and nature of computation allowed, fully homomorphic encryption (FHE) supports arbitrary computation on ciphertexts but at the expense of more substantial computing requirements.

◗ *Password-authenticated Key Agreement (PAKE)*. A PAKE protocol enables interacting parties to authenticate each other and derive a shared cryptographic key using a one or more party's knowledge of a password. The two parties prove possession of a shared password, but without revealing it or transmitting it over the insecure channel. As such, it protects against brute force dictionary attacks from eavesdroppers. [12]

◗ *Blockchain*. Blockchains offer an approach to implementing an immutable and distributed digital ledger, characteristically with no central repository and no central authority. Blockchains make extensive use of cryptographic hashing to create digests for transaction blocks, and public key cryptography to digitally sign them and to protect pseudonymity.

◗ *Threshold Cryptography*. In threshold cryptography, a private key is split and shared across multiple parties who are then able to reconstruct the key from a threshold number of participants who must cooperate to decrypt a message. The approach protects the secrecy of the private key and can be broadly applied: enabling digital signatures without a single party holding the entire signing key, performing encryption and decryption even if one component of the key becomes compromised, and more. [11]

One key open question is how PQC migration and cryptographic agility will apply to each approach. For example, how might cryptographic agility be added to blockchains, which are not designed for it? How will blockchain implementations navigate migration from RSA-based signatures, DSA, and ECDSA to PQC alternatives? One interesting observation is that many of the techniques are not yet widely deployed which means they can be agile with new features.

A second open question is whether any of these cryptographic technologies can help with the problem of cryptographic agility and PQC migration. For example, blockchain might be used to create a certificate ledger which manages certificates in use in an open, distributed, and secure way. Blockchain, and perhaps other schemes, might provide anchors of trust for both migration and agility challenges in various ways.

## 5. Conclusions

As quantum computing continues to make advancements in qubit technologies, scaling architectures, algorithms, applications, software tools, and more, it simultaneously fuels the urgency for a major transition in cryptography across the Internet as we know it today. At the time of this writing, NIST is actively leading an international community effort to select new public key cryptography algorithms for standardization to replace widely used RSA, ECDSA, ECDH, and DSA which are known to be vulnerable to attack by scaled quantum computing. Referred to as *post-quantum cryptography*, or *PQC*, these new algorithms use mathematical frameworks with no known mapping to quantum algorithms and are thus regarded as quantum resistant or quantum safe.

While NIST's much-needed standardization initiative addresses the selection of new cryptographic algorithms, *there is an urgent need to consider the complex problem of migration from today's widely deployed algorithms to PQC alternatives*. Underlying the explosive growth of the Internet and the "digitization" of nearly every sector of our society are security and privacy technologies that depend heavily on our current public key cryptography algorithms. The extensiveness of cryptography usage across the Internet means that an industry-wide migration to quantum safe cryptography standards (i.e., PQC) will be a massive undertaking, and one that is complicated by the layered complexity and heterogeneity of the worldwide compute infrastructure we operate. Prior history shows cryptographic migrations (e.g., 3DES to AES, MD5 to SHA1, SHA1 to SHA2, RSA to ECC) to take a decade or more before achieving broad success.

On January 31-February 1, 2019, the Computing Community Consortium (CCC) held a workshop in Washington, D.C. to discuss research challenges associated with PQC migration. Participants from three distinct yet related communities, *cryptographers*, contributing to the NIST PQC standards effort, *applied cryptographers* with expertise in creating cryptographic solutions and implementing cryptography in real-world settings, and *industry practitioners* with expertise in deploying cryptographic standards within products and compute infrastructures, came together to discuss research challenges surrounding *PQC migration* and *cryptographic agility*. Participants agreed that *challenges will require essential contributions by the research community* who is uniquely positioned to explore new approaches and advance our understanding through empirical rigor, analysis of fundamentals, investigation of complex tradeoffs, and more.

Based on workshop discussion, the recommendations are as follows:

*There is a need for research to understand and quantify the implications of replacing today's public cryptography algorithms with PQC alternatives across a wide variety of implementation and deployment contexts.* PQC candidate algorithms generally have greater computation, memory, storage, and communication requirements which imply the need to better understand performance, security, and implementation considerations. Research is needed on approaches to introducing new PQC algorithms (e.g., hybrids) within live systems that must remain interoperable with other systems during the period of industry migration. This includes such areas as formal modeling, automated tools, and approaching transition in complex infrastructures.

*While "cryptographic agility" is a familiar notion within the cryptography research community, there is a need to broaden and recast its scope in light of the size and complexity of global PQC migration.* A new *science of cryptographic agility* should include an expanded set of goals, a more comprehensive set of compute domains, a broader range of agility modalities and time scales, and the full range of security and computer science research methodologies. Research on cryptographic agility should include frameworks and architectures that enable agility across a wide variety of compute contexts, usable interfaces addressing various user roles, a better understanding of security and complexity tradeoffs, and other defining challenges. *Context agility*, or cryptographic frameworks that automatically select among algorithms and configuration based on the context of use, represents a long-term research vision that could shape the field.

*An important but overlooked area of research is social and policy aspects of cryptographic migration and agility.* In fact, policy, process, and people determine whether and when PQC adoption occurs at all. How should incentives be created for software developers and vendors to build cryptographic agility into their solution architectures? How might the role of government, customers, and the Internet marketplace contribute to creating the right incentives?

Finally, *research is also needed on the frontiers of cryptography;* that is, how PQC migration and cryptographic agility will apply to newer cryptography fields like secure multi-party computation, fully homomorphic encryption, blockchain, and more. This includes both how PQC migration and cryptographic agility will apply to each approach, and whether these developing cryptographic technologies can play a role in providing cryptographic agility and facilitating PQC migration.

## Appendix

Workshop Participants

| First Name | Affiliation |
| --- | --- |
| Reza Azarderakhsh | Florida Atlantic University |
| Shannon Beck | National Science Foundation (NSF) |
| Andy Bernat | Computing Research Association (CRA) |
| Matt Campagna | Amazon Web Services |
| Khari Douglas | Computing Community Consortium (CCC) |
| Ann Drobnis | Computing Community Consortium (CCC) |
| Roberta Faux | BlackHorse Solutions |
| Shay Gueron | University of Haifa, Amazon Web Services |
| Shai Halevi | IBM Research |
| Peter Harsha | Computing Research Association (CRA) |
| Mark Hill | University of Wisconsin (Madison), Computing Community Consortium (CCC) |
| Jeff Hoffstein | NTRU |
| David Jao | University of Waterloo |
| Sandip Kundu | National Science Foundation (NSF) |
| Hugo Krawczyk | IBM Research |
| Brian LaMacchia | Microsoft Research |
| Susan Landau | Tufts University |
| David McGrew | Cisco Systems |
| Ilya Mironov | Google Research |
| Rafael Misoczki | Intel Labs |
| Dustin Moody | National Institute of Standards and Technology (NIST) |
| David Ott | VMware Research |
| Kenny Paterson | ETH Zürich |
| Chris Peikert | University of Michigan (Ann Arbor) |
| Radia Perlman | Dell EMC |
| Tom Ristenpart | Cornell Tech |
| Vladimir Soukharev | InfoSec Global |
| Helen Wright | Computing Community Consortium (CCC) |
| Rebecca Wright | Barnard College of Columbia University |

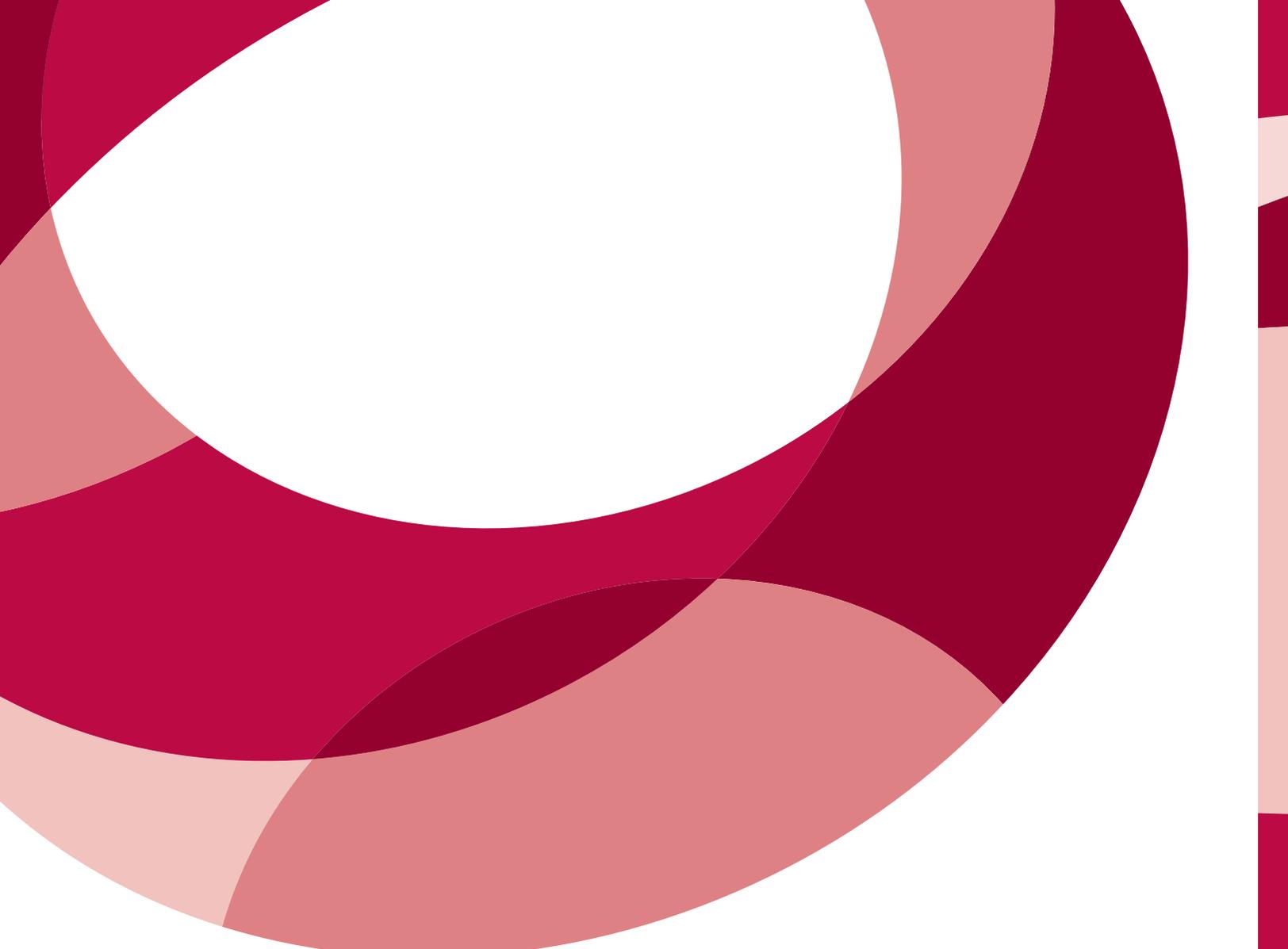

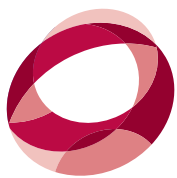

**CCC**
Computing Community Consortium
Catalyst

1828 L Street, NW, Suite 800
Washington, DC 20036
P: 202 234 2111 F: 202 667 1066
www.cra.org cccinfo@cra.org